%% file: main.tex
\documentclass[conference, a4paper]{IEEEtran}
\usepackage[top=.75in, left= 0.625in, right = 0.63in, bottom =1in]{geometry}
\usepackage{amsmath,amssymb,amsthm,algorithm,algpseudocode,flushend}
\usepackage{array}
\usepackage{fixltx2e}
\usepackage{stfloats}
\usepackage[utf8]{inputenc}	
\usepackage[british]{babel}
\usepackage{csquotes}
\usepackage[pdftex]{graphicx}
\graphicspath{ {image/} }
\usepackage{mathtools}
\usepackage{todonotes}
\usepackage{authblk}

\begin{document}
\title{Time Reversal for Multiple Access and Mobility: Algorithmic Design and Experimental Results}

\author[1]{Ali Mokh}
\author[1]{Julien de Rosny}
\author[2]{George C. Alexandropoulos}
\author[3]{\\Ramin Khayatzadeh}
\author[3]{Mohamed Kamoun}
\author[1]{Abdelwaheb Ourir}
\author[1]{Arnaud Tourin}
\author[1]{Mathias Fink}
\affil[1]{ESPCI Paris, PSL Research University, CNRS, Institut Langevin, France}
\affil[2]{Department of Informatics and Telecommunications,
National and Kapodistrian University of Athens, Greece}
\affil[3]{Mathematical and Algorithmic Science Lab, Paris Research Center, Huawei Technologies France}
\affil[ ]{emails: firstname.lastname@espci.fr, alexandg@di.uoa.gr, firstname.lastname@huawei.com}

\maketitle

\begin{abstract}
Time Reversal (TR) has been proposed as a competitive precoding strategy for low-complexity wireless devices relying on Ultra-WideBand (UWB) signal waveforms. However, when TR is applied for multiple access, the signals received by the multiple users suffer from significant levels of inter-symbol and inter-user interference, which requires additional processing for mitigation by each receiving user. In this paper, we present an iterative Time-Reversal Division Multiple Access (TRDMA) approach that aims to dim the latter interference levels. The performance of iterative TRDMA is evaluated experimentally in a reverberation chamber that mimics a rich scattering indoor wireless propagation environment. The improved efficiency, in terms of the number of algorithmic iterations, of the proposed approach compared to conventional TRDMA, is demonstrated. We also consider a mobile user configuration, where the position of the receiver changes between the channel estimation and data transmission steps. It is showcased, even for this experimental setup, that the proposed iterative TRDMA approach is more efficient than conventional precoding schemes.
\end{abstract}


\IEEEpeerreviewmaketitle

\input{Introduction}
%
\input{Theory}

\input{Experimental}
%

\section{Conclusion}
In this paper, we have presented indoor experimental results for an UWB multiple access scheme based on the TR technique. We have proposed a novel iterative algorithm to confront the effects of ISI and IUI in TR-precoded wireless communication systems. The efficiency of the presented ITRDMA scheme has been validated through real experiments in a highly scattering medium. The presented approach includes a multi-user MISO channel sounding process to estimate the CIRs, which was used to derive the precoding scheme. Then, the TR-precoded received intended signal as well as the interference were measured in baseband and used for the performance assessment. Our performance results showcased that the proposed ITRDMA algorithm is capable to reduce the interference in the received signal, and therefore increase the SINR level. We also investigated the effect of mobility of a receiving user using the same ITRDMA scheme. It was demonstrated that the proposed precoding scheme can still be more efficient than conventional TR at high SNRs, and up to moderate speed values that are reasonable for indoor wireless communications. In future works, we will investigate the performance of the ITRDMA algorithm in imperfect channel estimation settings, and will perform comparisons with other spatial precoding techniques (e.g., regularised zero-forcing). We also intend to experiment data communication considering reconfigurable intelligent surfaces \cite{alexandg_2021} for reprogrammable rich scattering conditions.
\bibliographystyle{IEEEtran}
\bibliography{IEEEabrv,ref}
\end{document}

%% file: Introduction.tex
\section{Introduction}
Ultra-WideBand (UWB) signal waveforms have ushered a new era in short-range data transmissions for wireless sensor networks, by enabling both robust communications and accurate ranging capabilities \cite{zhang2009uwb}. One central issue facing UWB waveforms for low complexity devices refers to effective energy collection that is usually dispersed in rich scattering wireless environments. The Time Reversal (TR) technique appears to be a paradigm shift in exploiting rich multipath signal propagation, which is capable of focusing UWB pulses both in time and space \cite{lerosey2004time}, while shifting the signal processing complexity to the transmitter side.

With TR precoding, the receivers can benefit from high levels of the intended received signals enabling, in this way, simple, low cost, and low power reception modules \cite{wang2011green,chen2013time}. Moreover, the offered temporal pulse compression and spatial focusing can partially mitigate the Inter-Symbol Interference and the Inter-User Interference (IUI), resulting from the multipath signal propagation \cite{oestges2004time}. The above characteristics render TR precoding a promising scheme for wireless communications and localization applications. Very recently, \cite{Alexandropoulos_ICASSP} demonstrated indoor cm-level localization accuracy with TR at $3.5$ GHz using up to $600$MHz bandwidth channel sounding signals. In \cite{Mokh_CSCN}, considering a leaking reverberation cavity, the authors experimentally verified the ultra-accurate spatiotemporal focusing capability of TR at $273.6$ GHz using transmission bandwidths up to $4$ GHz.

In \cite{han2012time}, a TR-based Division Multiple Access (TRDMA) scheme for multi-user Multi-Input Single-Output (MISO) systems was presented. It was shown that the proposed scheme provides a cost-effective waveform design enabling energy-efficient Internet-of-Things (IoT) nodes \cite{wang2011green}. Thanks to this simple precoding strategy, base stations can support a large number of simultaneous links with moderate complexity \cite{han2014multiuser}. A theoretical analysis for the average Signal-to-Interference Ratio (SIR) of the intended and unintended receivers in a distributed TR-based transmission scheme was carried out in \cite{wang2016snr}, while an experimental evaluation has been recently presented in \cite{mokh2021indoor}. However, it is known that when operating in low Signal-to-Noise Ratio (SNR) values, the interference level can affect the TR performance. Indeed, one major issue with TRDMA is that the SIR level cannot be tuned. For instance, it has been shown that in an ideal Rayleigh channel, the averaged SIR is equal to $0$ dB. 

To overcome the latter problem with TR while keeping the receiver as simple as possible, the iterative TRDMA (ITRDMA) technique was proposed in \cite{montaldo2004telecommunication}, aiming to reduce the ISI and IUI. This technique belongs to the family of pre-rake processing techniques, which were originally introduced in \cite{imada2005pre}. It is noted that iterative processing is a popular approach for equalization. It has been applied for single and multi-carrier code division multiple access (CDMA) systems \cite{liu2004iterative}. Very recently, a low complexity iterative rake detector for orthogonal time-frequency space modulation has been presented in \cite{thaj2020low,thaj2020low2}. Although the ITRDMA scheme proposed in \cite{montaldo2004telecommunication} deals with ISI and IUI, it is quite difficult to implement. It requires multiple feedback between the transmitters and the receivers, which makes it non-applicable for very low duty cycle nodes. 

Motivated by the latter limitation, in this paper, we present an ITRDMA algorithm that is based on the iterative processing of the Channel Impulse Responses (CIRs) of the multiple users, which are acquired during the channel estimation phase. The proposed algorithm is implemented at the transmitter side as a precoder, instead of being implemented at the receiver as a post-processing step. Hence, the ITRDMA computational complexity is concentrated at the transmitter (e.g., the base station or access point) instead of the receiver, when downlink communication is considered. The performance of the proposed algorithm, in terms of ISI and IUI mitigation, is experimentally evaluated in a multi-user scenario inside a reverberating room mimicking a strongly scattering indoor environment. We specifically validate the performance using channels that have been experimentally acquired at the carrier frequency $2$ GHz using an $100$ MHz communication bandwidth. It is showcased, even for an experimental setup with mobility, that the proposed ITRDMA approach is more efficient than conventional TRDMA schemes.


%% file: Theory.tex
\section{Time Reversal for Multi-User Precoding}
\label{SEC:TRDMA}
In this section, we first introduce the TR precoding and then present the proposed ITR scheme for multi-user wireless systems. 
%
\subsection{System Model and Time Reversal Precoding}
We consider a multi-user MISO wireless communication system comprising one base station that is equipped with $M$ antenna elements, and which wishes to communicate in the downlink direction with $N$ single-antenna users. We represent by $h_{i,m}[k]$ the baseband CIR at discrete time $k$ between the $m$-th ($m=1,2,\ldots,M$) transmitting and $i$-th ($i=1,2,\ldots,N$) receiving antennas, i.e., at the base station and the $i$-th user, respectively. The TR precoding technique utilizes the time-reversed CIR, i.e., $h^*_{i,m}[L-k]$, to focus the transmitted information-bearing electromagnetic field on the $i$-th user location. As a consequence, the signal sent by the $m$-th transmit antenna at discrete time $k$ to focus the baseband message $x_i$ on each individual receiving antenna is given by:
\begin{align}
     s_m[k]=\sum_{i=1}^N  \frac{\sum_{l=1}^L x_i[l] h_{i,m}^*[L+l-k]}{\sqrt{\sum_{m=1}^M\sum_{l=1}^L \lvert h_{i,m}[l] \rvert^2}}, 
     \label{eq:yi}
\end{align}
where it is assumed that the CIR is composed of $L$ significant taps. The normalization in this expression ensures that the power emitted toward each user is the same. Using \eqref{eq:yi}, the baseband received signal at each $j$-th ($j=1,2,\ldots,N$) user at time $k$ can be expressed as follows:
\begin{equation}
   y_j[k]= \sum_{l=1}^L \sum_{i=1}^N x_i[l] R_{j,i}[L+l-k]+n_j[k].
   \label{eq:TR}
\end{equation}
where the expression of the correlation function $R_{j,i}[k]$ between the CIRs of users $j$ and $i$ is defined as 
\begin{equation}
 	R_{j,i}[k] \triangleq \frac{\sum_{m=1}^M\sum_{k'=-\infty}^{+\infty} h_{i,m}^*[k-k'] h_{j,m}[k']}{\sqrt{\sum_{m=1}^M\sum_{l=1}^L \lvert h_{i,m}[l] \rvert^2}},
   \label{eq1}
\end{equation}
and $n_j$ represents the zero-mean additive white Gaussian noise with standard deviation $\sigma$ at user $j$. 

In multipath channels, when applying TR precoding using the CIR of user $i$, the time-reversed field focuses in time and space in the location of this user \cite{fink2001acoustic}. This means that a signal peak given by the CIR's auto-correlation function $R_{i, i}[0]=\sqrt{\sum_{m'=1}^M\sum_{l=1}^L \lvert h_{i,m}[l] \rvert^2}$ appears at user $i$; at this time and location, $ML$ signals add up coherently. However, this term can be corrupted by incoherent contributions of the auto-correlation function $R_{i,j}[k]$. In fact, the contributions when $k\neq0$ but $i=j$ are due to imperfect time focusing on the targeted user/antenna. This implies that around the main focused peak, some secondary peaks are still present that generate ISI. Note that the elements of the auto-correlation function when $i \neq j$ happen due to the sides lobes. Hence, TR precoding does not perfectly focus the wave on the intended user, inducing interference to the other users.

\subsection{Proposed Time-Reversal-Based Multiple Access}
To reduce the levels of ISI and IUI appearing in \eqref{eq1}, we propose the ITRDMA algorithm. In contrary to the scheme presented in \cite{montaldo2004telecommunication}, our algorithm is based on the measurement of the CIRs of the multiple users. Its main principle is to iteratively remove the secondary peaks and lobes in the autocorrelation function. To this end, the time position and the antenna location of the targeted lobe are identified iteratively. In particular, a pre-processing time-reversed waveform is used to focus on that position at that time index, but with a peak that is the opposite of the complex amplitude of the lobe. Hence, if this time-reversed waveform is added to the actual CIR, thanks to the linearity, the secondary lobe will be canceled. The same procedure can be used iteratively to remove all secondary peaks and lobes from the autocorrelation function. The algorithmic steps of the proposed ITRDMA scheme are summarized as follows:
\begin{enumerate}
    \item The transmit precoder ${s}_{i_0,m}[k]$ at the discrete time $k$ is initialized with $\tilde{h}^*_{i_0,m}[L-k]$ to focus the received signal on the user/antenna $i_0$ at the time instant (i.e., channel tap) $k=L$. The notation $\tilde h$ stands for the normalized CIR, which is computed as follows:
    \begin{equation}
   \tilde h_{i,m}[k] \triangleq  h_{i,m}[k]\left(\sqrt{\sum_{m=1}^M\sum_{k'=1}^{L} \lvert h_{i,m}[k'] \rvert^2}\right)^{-1}.
   \label{eq:Normalized_CIR}
\end{equation}
	\item The differences $\Delta_i[k]$ $\forall$$i=1,2,\ldots,N$ between the resulting fields $f_i[k]$ on the $N$ receiving users/antennas, which are given by the autocorrelations $\tilde{R}_{i_0,i}[L-k]$ and their targeted values $\delta[L-k]\delta[i-i_0]$ (i.e., for the fields without lobes), are computed for each receiving antenna. The notation $\tilde{R}_{j,i}[k]$ with $i,j=1,2,\ldots,N$ represents the autocorrelation function using the normalized CIRs, which is given by the expression: 
	\begin{equation}
   \tilde{R}_{j,i}[k] = \sum_{m=1}^M\sum_{k'=-\infty}^{+\infty}  \tilde{h}_{i,m}^*[k-k'] \tilde{h}_{j,m}[k'].
   \label{eq:Normalized_auto}
\end{equation}
	\item The tap index $\hat{k}$ and the antenna index $\hat{i}$ where the maximum of $\lvert \Delta_i[k] \rvert$ occurs are found. This corresponds to the highest (undesired) secondary peak in the autocorrelation function.
	\item The ITRDMA precoder is then updated with the time-reversed CIRs, which are shifted in time to focus on the $\hat{i}$-th user/antenna at the $\hat{k}$-th tap with a complex amplitude opposite in sign with $ \Delta_{\hat{i}}[\hat{k}]$, in order to cancel the secondary peaks on each receiving user.
	\item The differences $\Delta_i[k]$ $\forall$$i$ are also updated to take into account the new contributions to the precoder field. it will hold that $\Delta_{\hat{i}}[\hat{k}]=0$.
\end{enumerate}

The algorithmic steps of the proposed ITRDMA scheme are summarized in Algortihm~\ref{Hybrid_F}. It can be seen that the iterative Steps $6$ to $8$ remove the secondary taps via TR-based precoding. The iterations stop either when their maximum allowable number $n_{\max}$ is reached or when $\left|\Delta_{i}[k]\right|$ becomes smaller than a target value $\epsilon$. It is noted that this value can be advantageously tuned to meet a target SNR level or to tune the complexity and the processing latency at the transmitter side. Finally, at the Step $11$ of the algorithm, the TR-based precoding signals $s_{i,m}[k]$ at each time discrete index $k$ are normalized in energy. 
%
\begin{algorithm}[!t]
\caption{The Proposed ITRDMA Precoding Scheme}\label{Hybrid_F}
\textbf{Initialize:} Set the values for $M$, $N$, and $L$.
\begin{algorithmic}[1]
\For {$i_0=1,2,\ldots,N$}
\State Set ${s}_{i_0,m}[k] = \tilde{h}^*_{i_0,m}[L-k]$.
\State Set $\Delta_i[k]  = \tilde{R}_{i_0,i}[L-k] - \delta[L-k]\delta[i_0-i]$.
\State Set $n=0$.
\While {$\max \lvert \Delta_i[k]\rvert>\epsilon$ and $n<n_{\max}$} 
\State Find $\hat{k},\hat{i} = \arg \max_{k,i} \lvert \Delta_i[k]\rvert$. 
\State Set ${s}_{i_0,m}[k] \leftarrow {s}_{i_0,m}[k]-\Delta_{\hat{i}}[\hat{k}] \tilde{h}^*_{\hat{i},m}[L-k+\hat{k}]  $.
\State Set $\Delta_i[k] \leftarrow \Delta_i[k]-\Delta_{\hat{i}}[\hat{k}] \tilde{R}_{\hat{i},i}[L-k+\hat{k}]$.
\State Set $n=n+1$.
\EndWhile
\State Set $s_{i_0,m}[k] \leftarrow {s}_{i_0,m}[k]\left(\sqrt{\sum_{k',m}|{s}_{i_0,m}[k']|^2}\right)^{-1}$.
\EndFor
\end{algorithmic}
\end{algorithm}

\subsection{Signal-to-Interference-Plus-Noise (SINR) Performance}
The complex-valued information symbols $x_i[l]$ intended for each $i$-th user are convoluted with the precoding sequences $s_{i,m}[k]$ computed from Algortihm~\ref{Hybrid_F}. As a consequence, the baseband received signal at each $j$-th user can be expressed as follows:
\begin{equation}
y_j[k] = \sum_{l=1}^{L}\sum_{i=1}^{N} x_j[l] w_{i,j}[k-l] +n_j[k],
\end{equation}
where $w_{i,j}[k]$ represents the precoding signal given by
\begin{equation}
w_{i,j}[k] \triangleq \sum_{l'=1}^L\sum_{m=1}^M s_{j,m}[l'] h_{i,m}[k-l'].
\end{equation}
In the case of TR precoding, by inspection of expression (2), it can be concluded that $w_{i,j}[k] = R_{i,j}[L-k]$. Therefore, the sequence $w_{i,j}[k]$ can be interpreted as a virtual channel that includes the TR-based precoding. This equivalent channel approach has been already use for TR in \cite{phan2013make}. By assuming that the power of the information symbols is normalized, it is easy to show that SINR of the equivalent channel at each receiving user/antenna $i$ is obtained as follows:
\begin{equation}
{\rm SINR}_i \triangleq \frac{|w_{i,i}[0]|^2}{I_i + \sigma^2}.
\label{eq:SINR}
\end{equation}
where $I_i\triangleq\sum_{l=1,l\neq 0}^L |w_{i,i}[l]|^2+ \sum_{l=1}^L\sum_{j=1,j\neq i}^N |w_{j,i}[l]|^2$ represents the intereference term.

%% file: Experimental.tex
\section{Experimental Validation Results}

\subsection{Experimental Setup}
\begin{figure}[t!]
\centering
\includegraphics[width=\linewidth]{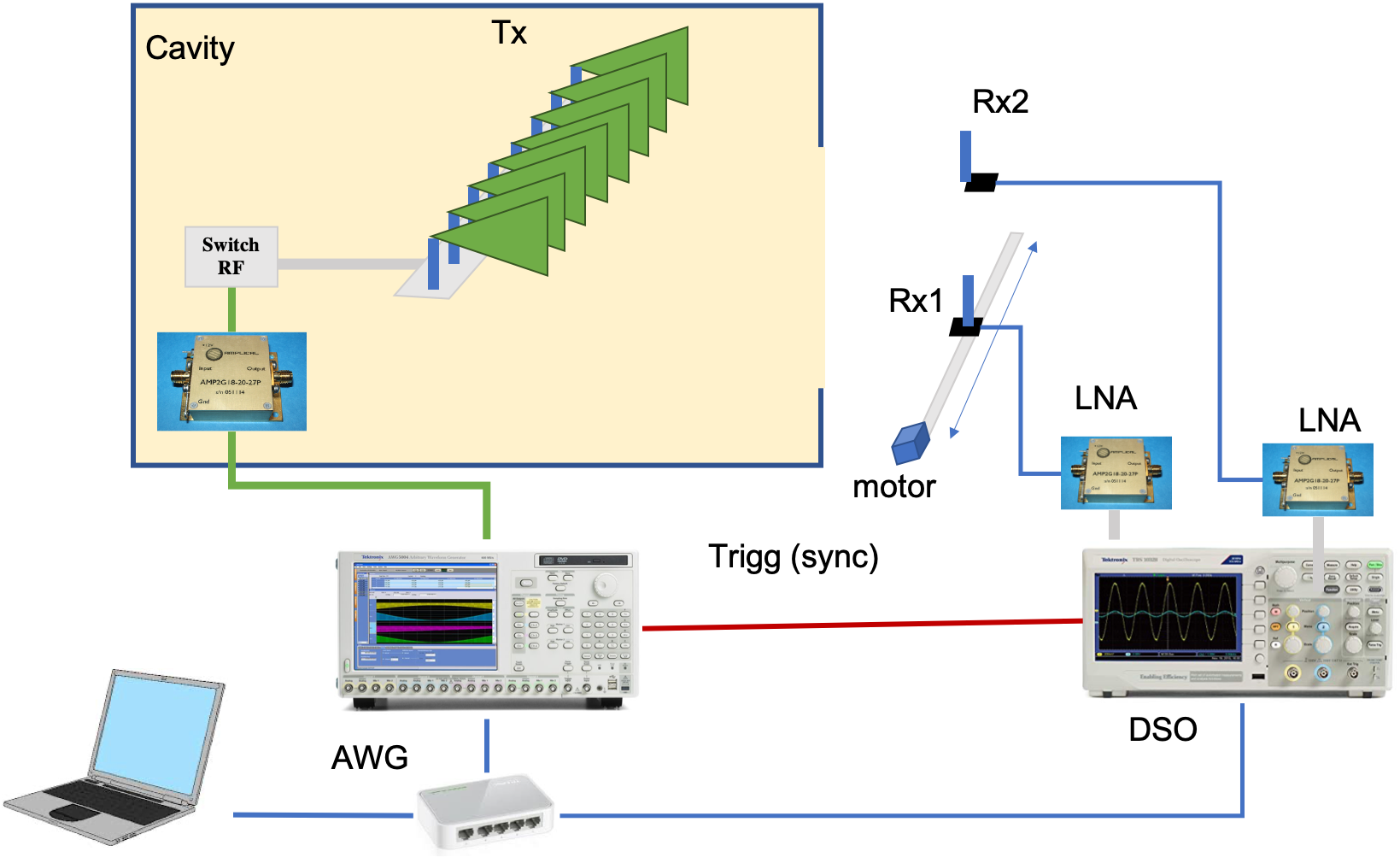}
\caption{The designed and realized experimental setup for the proposed TR-based multi-user MISO scheme.}
\label{fig:setup} 
\end{figure}
To evaluate the performance of the proposed ITRDMA precoding scheme, we have implemented an experimental setup to first sound the channel and then to TR-precode the transmitted signals, as shown in Fig.~\ref{fig:setup}. The setup consists of a transmitter array with $M=8$ antennas, all connected to a solid-state Radio Frequency (RF) switch. The input of this $1\times8$ switch is connected to a power amplifier that is fed by an Arbitrary Waveform Generator (AWG) from Tektronix (specifically, the AWG-7012). This AWG was used to generate transmitted signals with a carrier frequency at $2$ GHz. The transmitter (Tx) was placed inside a reverberating cavity to create a rich multipath signal propagation channel. This cavity has an opened window facing the Receivers (i.e., Rx1 and Rx2), which are placed outside the cavity. The AWG and DSO were connected to an Ethernet network switch and controlled by a desktop computer. The AWG triggered the DSO to ensure synchronization. The sampling rate of the DSO and the AWG was $12.5$ and $10$ GHz, respectively.

Two different setup configurations were used for two different sets of measurements. The first setup intended to evaluate the ITRDMA precoding scheme in a two-user MISO scenario, where two dipole antennas were connected to two low noise amplifiers, and then, to the inputs of an RF Digital Storage Oscilloscope (DSO) (specifically, the Tektronix TDS6604B). The second setup was used to evaluate the TR precoding in a mobile setting, considering a single-user MISO scenario. In particular, the receiving user Rx1 was mounted on a linear motorized bench in order to be able to move its receiving antenna without strongly modifying the channel property inside the reverberating cavity. The bench was placed parallel to the cavity's opening, resulting in a displacement range equal to $30$cm.

\subsection{Spatiotemporal Focusing Capability}
\begin{figure}[t!]
\centering
\includegraphics[width=1 \linewidth]{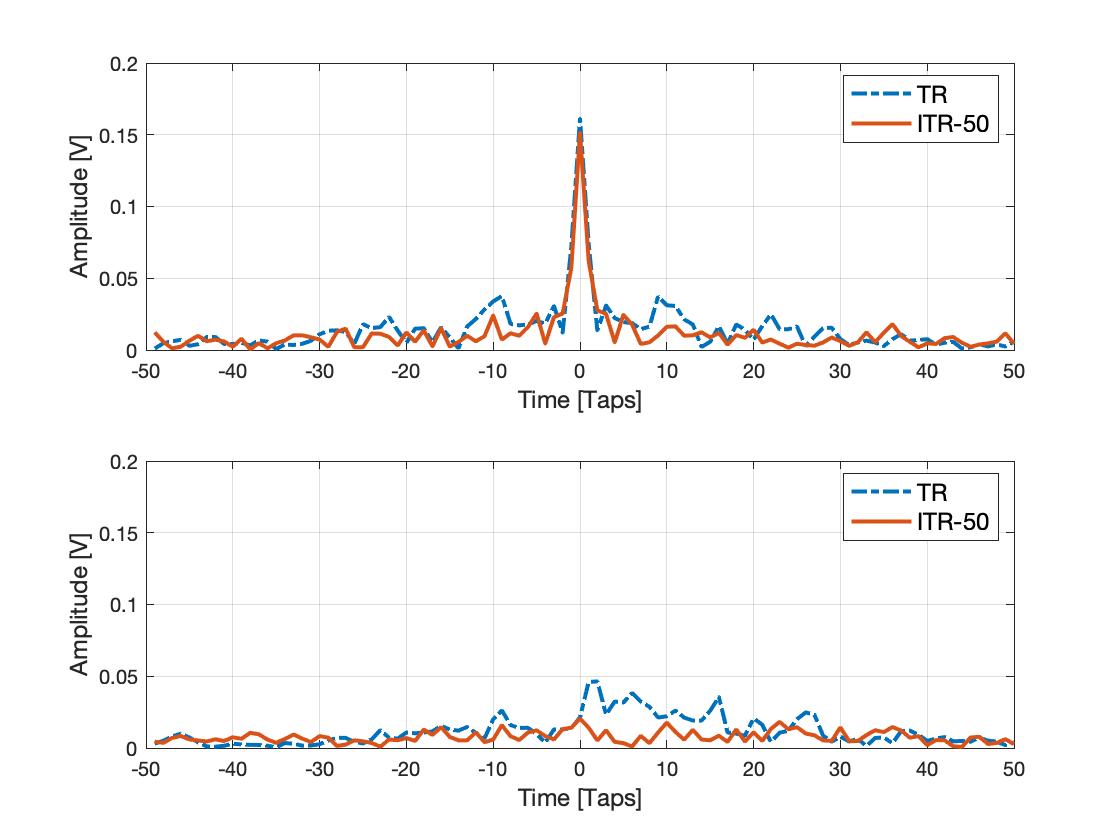}
\caption{Amplitude in Volts of the baseband received field probed by the antenna of Rx1 (top figure) and the antenna of Rx2 (bottom figure), when TR and ITRDMA precoding are considered by the $8$-antenna transmitter array in order to spatiotemporally focus the transmitted signal at Rx1.}
\label{fig:TRVsITR} 
\end{figure}

\begin{figure}[t!]
\centering
\includegraphics[width=1 \linewidth]{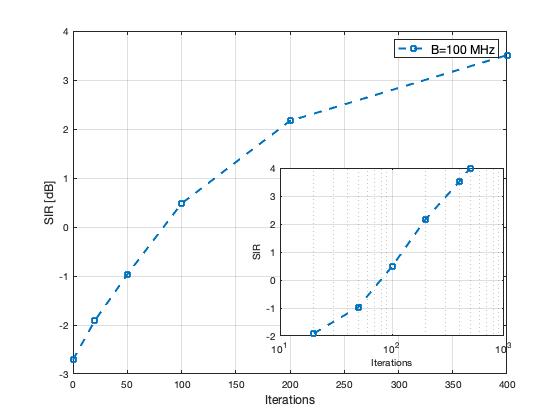}
\caption{The SIR performance of the proposed ITRDMA precoding algorithm with respect to the number of iterations using $B=100$ MHz. The inner plot is the same plot, but its horizontal axis is in log scale.}
\label{fig:SINRtoiter} 
\end{figure}

In the first experiment, we ran the transmission chain with a bandwidth $B=100$ MHz to focus the precoded signal on the user Rx1. We applied both the conventional TR and the proposed ITR precoding algorithms using $50$ iterations. The radiated field was probed not only on the user Rx1, but also on the user Rx2. The amplitude of the baseband received fields versus the discrete time index is demonstrated in Fig.~\ref{fig:TRVsITR}. We first clearly observe a strong time and spatial focusing with the both TR-based precoding algorithms. However, the side lobes on Rx1 and the secondary lobes on Rx2 are smaller when ITRDMA is applied. One can also see that the ITRDMA-precoded signals have slightly lower amplitude at time $t=0$.
To quantify the improvement of the ITRDMA-based focusing, we plot in Fig.~\ref{fig:SINRtoiter} the SIR performance of the received signal, as computed via \eqref{eq:SINR} by setting $\sigma=0$, as a function of the number of the algorithmic iterations for a $B=100$ MHz bandwidth signal. Note that the SIR of conventional TR precoding is obtained when the number of iterations with ITR equals to $0$. It can be seen that, thanks to the proposed iterative scheme, an improvement up to $6.5$ dB is obtained with ITR for $400$ iterations. Evidently, the improvement is not linear with the number of iterations. We actually observe that the SIR improves with the square root of this number, as soon as the iterations are more than $50$. We believe that those improvement and behavior strongly depend on the channel characteristics and transmission bandwidth.

\subsection{The Effect of Mobility}
\begin{figure}[t!]
\centering
\includegraphics[width=1 \linewidth]{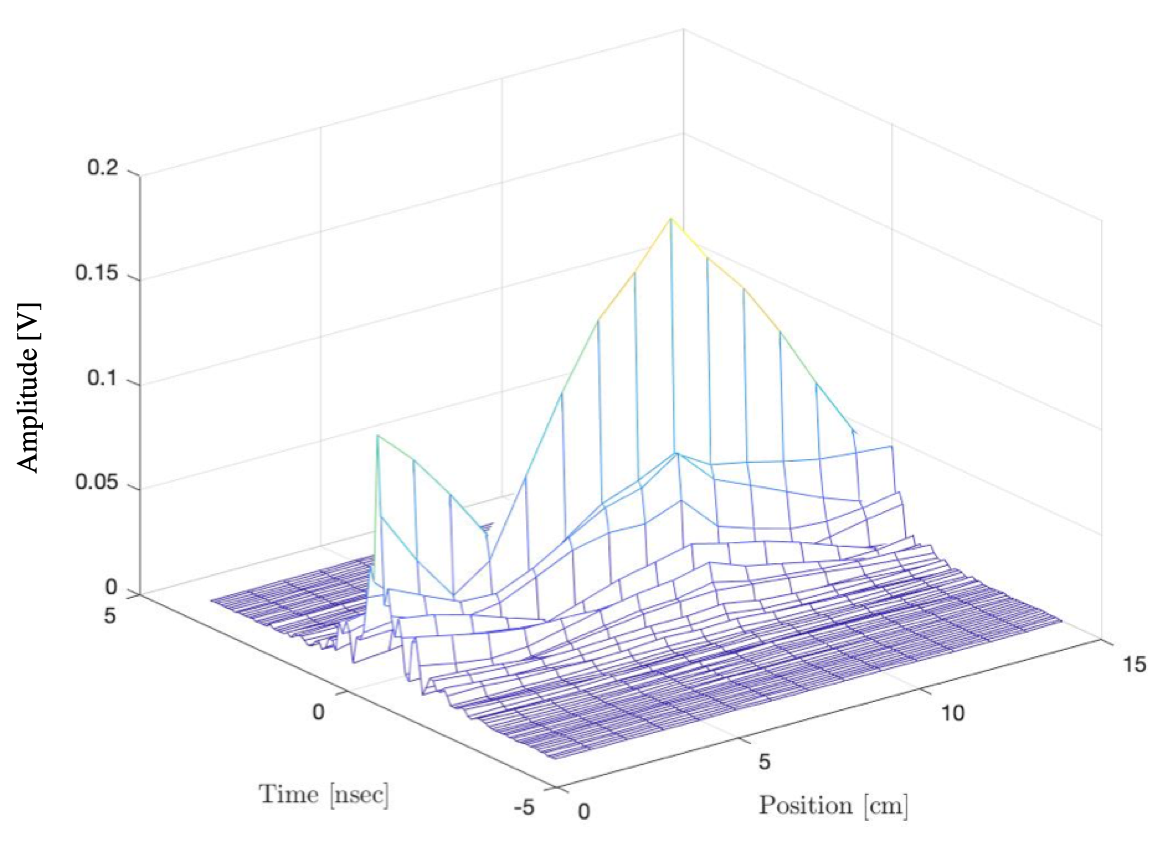}
\caption{Amplitude [V] of the baseband received signal with respect to time and the position of the antenna of Rx1. The ITRDMA-precoded signals have been obtained using $50$ algorithmic iterations and the transmission bandwidth $B=100$ MHz.}
\label{fig:SPfocus} 
\end{figure}

The TR-based precoding is known to be sub-optimal, especially in the case of low-to-moderate noise levels. However, it is also known to be less sensitive to channel fluctuations and receiver mobility. Hence, one may wonder to what extent the proposed ITRDMA processing is more sensitive than conventional TR, when considering a displacement of a receiving user between its position when channel estimation was performed and its new position during the data transmission phase. We experimentally investigate this question, as follows. We consider the user Rx1 lying on the motorized bench, and a first step, the $8$ CIRs are estimated for an initial position of Rx1 at the point $x=8$ cm. Using those estimates, the ITRDMA precoder is derived. In a second step, the Rx1 is moved between the positions $x=1$ and $15$ cm with a step size of $1$ cm. For each of those positions, the same precoder was applied and the measured signal in baseband was recorded and stored. This signal's amplitude, considering $50$ algorithmic iterations for ITRDMA, is illustrated as a function of time and RX1 position in Fig.~\ref{fig:SPfocus}. As expected the maximum of the spatiotemporal focusing is obtained for the Rx1 position at $x=8$ cm, where the CIR was estimated. It is also shown that the maximum amplitude reaches a minimum when the receive antenna is moved by $5$ cm from the CIR measurement position. For TR  preprocessing, it is well known that the width of the focal spot is directly given by the spatial coherence length, which is equal to half a wavelength for a diffuse field. In our experiment at $2$ GHz carrier frequency, this coherence length is $3.4$ cm. The larger value seen in this figure happens because the receiving antenna is placed outside the reverberation cavity, leading to an increase of the spatial coherence length.

\begin{figure}[t!]
\centering
\includegraphics[width=1 \linewidth]{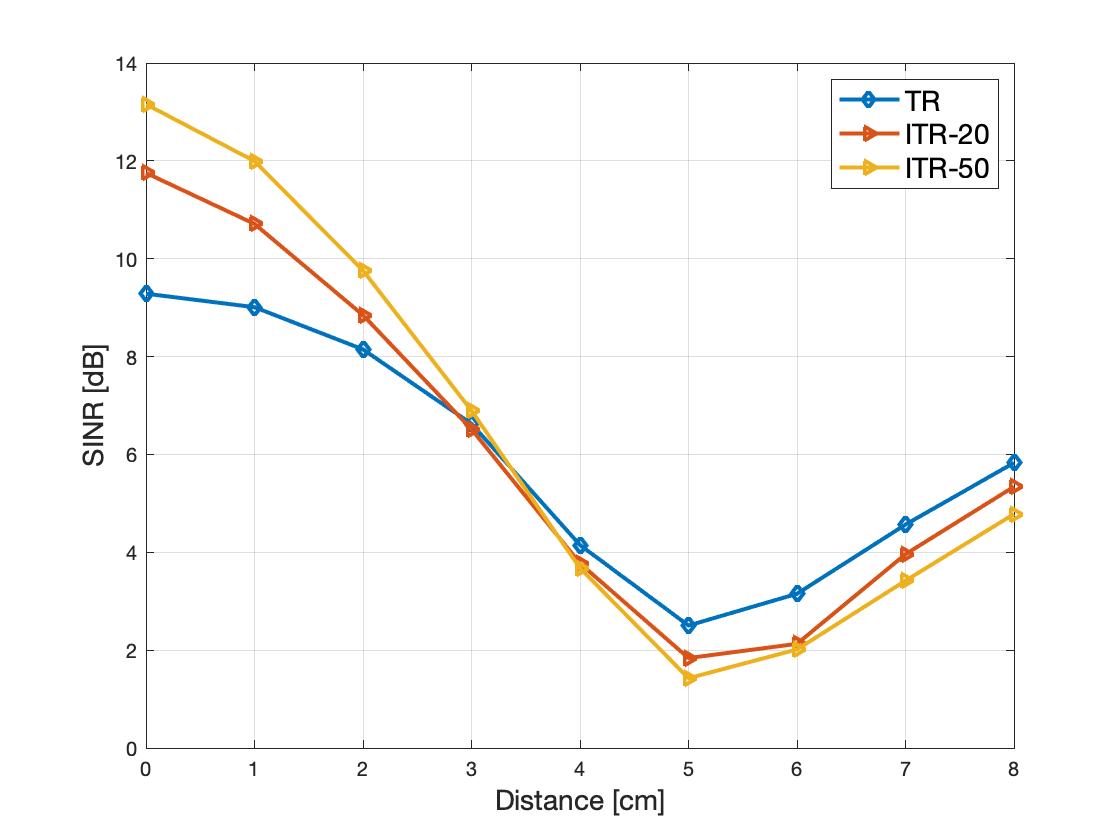}
\caption{The estimated SINR at Rx1 with respect to the distance traveled by this user between the CIR estimation process and the application of the ITRDMA precoder. The performance curves were obtained for $3$ different precoders: \textit{i}) conventional TR, as well as ITRDMA with \textit{ii}) $20$ and \textit{iii}) $50$ algorithmic iterations. The transmission bandwidth and operating SNR were set to $100$ MHz and $30$ dB, respectively.}
\label{fig:CH10SNR30} 
\end{figure}

We next investigate the effect of the user receiver displacement between the channel sounding and the data transmission phases in the SINR performance, by evaluating the expression \eqref{eq:SINR}. The Doppler effect was not taken into account. However, it is expected to have a moderate impact on the data transmission since we perform wideband processing in the time domain. The SINR curves versus the Rx1 displacement in cm for the SNR value $30$ dB are demonstrated in Fig.~\ref{fig:CH10SNR30}. We clearly observe that, for small distances, the larger the number of iterations is, the faster is the decay of the SINR with respect to the distance. Nevertheless, the ITRDMA algorithm seems to inherit the TR robustness. Indeed, once the SINR improvement due to the $50$ algorithmic iterations is lost, its behavior becomes identical to the one obtained with only $20$ iterations. Hence, by optimizing the number of iterations in our algorithm, one can get better performance with the proposed ITRDMA scheme than using conventional TR.

We hereinafter set as $\tau$ the time duration between the channel estimation process and the actual data transmission, ans as $v$ the speed of mobility of the user Rx1. Based on the findings in Fig.~\ref{fig:SPfocus}, one can assess the maximum speed that Rx1 can reach before its SINR performance reduces by a factor of $2$ for different CIR estimation time intervals $\tau$. Indicative results for this metric are provided in Table~I. The shown time duration values are completely compatible with our indoor applications of the ITRDMA algorithm. For outdoor applications, the ITRDMA processing has the same limitations with any precoding algorithm, where the channel needs to be estimated by the transmitter. Note that the speed values listed in Table~I are probably overestimated, because the outdoor coherence length can be larger. In this table, we have calculated the minimum speed of the user Rx1 in order to reach half of the signal strength for the corresponding time duration $\tau$. As clearly shown in the figure, a lower value of $\tau$ will result in SINR being more robust to mobility. 
\begin {table}[t!]
\caption{The Required Speed of User Rx1 to Reach Half of the Signal Strength For Different Channel Estimation Durations.}
\label{tab:speedtotau}
\begin{center}
\begin{tabular}{|c||c|c|c|}
\hline
$\tau$ (msec) & $50$ & $10$ & $1$ \\
\hline
$v$ (km/h) & $2.16$ & $10.8$ & $108$ \\
\hline
\end{tabular}
\end{center}
\end {table}

\begin{figure}[t!]
\centering
\includegraphics[width=1 \linewidth]{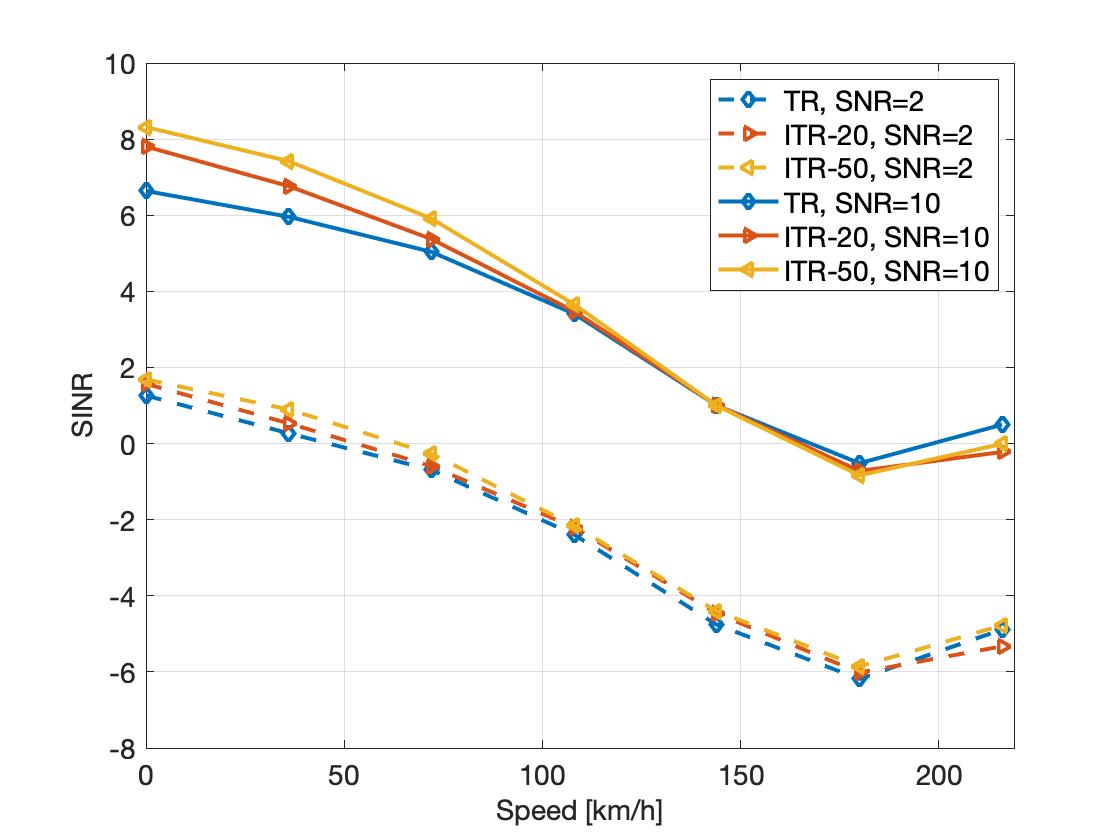}
\caption{The SINR performance versus the speed of the mobile user Rx1 lying in the motorized bench for transmission bandwidth $B=100$ MHz, time mismatch duration $\tau$=1 ms, and operating SNR equal to $2$ and $10$ dB.}
\label{fig:CH1SNR210} 
\end{figure}

Finally, in the SINR performance evaluation results in Fig.~\ref{fig:CH1SNR210}, we investigate the effect of the noise on the behavior of the different TR precoding schemes, when considering varying mobility speeds for the user Rx1. In this figure, the time duration $\tau$ between the CIR estimation and data transmission is set to $1$ ms, and we plot the SINR values of the baseband received signal when applying conventional TR, as well as ITRDMA with $20$ and $50$ iterations. Performance curves for operating SNR values equal to $2$ and $10$ dB are depicted. It is shown that, the lower the SNR is, the closer are the performances of the different TR-based precoding schemes. This is due to the fact that for high noise levels, the interference becomes lower than the noise level, and then the interference cancellation process becomes useless. For low-to-moderate user speeds, which are reasonable for indoor applications, it is demonstrated in the figure that ITRDMA outperforms conventional TR. However, for very large speeds, conventional TR, which is obtained from ITRDMA for zero iterations, is sufficient.